\documentclass[conference]{IEEEtran}
\usepackage{url}
\usepackage{cite}
\usepackage{amsmath,amssymb,amsfonts}
\usepackage{algorithmic}
\usepackage{graphicx}
\usepackage{subfigure}
\usepackage{textcomp}
\usepackage{xcolor}
\def\BibTeX{{\rm B\kern-.05em{\sc i\kern-.025em b}\kern-.08em
    T\kern-.1667em\lower.7ex\hbox{E}\kern-.125emX}}
    
\usepackage{amsthm}
\usepackage[ruled]{algorithm2e}
\usepackage{mathrsfs}
\usepackage{bm,epsfig,amsthm,url}
\usepackage{indentfirst}
\usepackage{epstopdf}

\theoremstyle{definition}
\newtheorem{lemma}{Lemma}

\begin{document}
	
%\title{Outage Minimization for Intelligent Reflecting Surface Aided MISO Communication}
\title{Outage Minimization for Intelligent Reflecting Surface Aided MISO Communication Systems via Stochastic Beamforming}
\author{
  \IEEEauthorblockN{Wenzhi Fang, Min Fu, Yuanming Shi, and Yong Zhou}
  \IEEEauthorblockA{School of Information Science and Technology,                    ShanghaiTech University, Shanghai 201210, China\\
					E-mail: wenzhi.fang@foxmail.com, \{fumin, shiym, zhouyong\}@shanghaitech.edu.cn
					}
}

\maketitle

\begin{abstract}
Intelligent reflecting surface (IRS) has the potential to significantly enhance the network performance by reconfiguring the wireless propagation environments. 
It is however difficult to obtain the accurate downlink channel state information (CSI) for efficient beamforming design in IRS-aided wireless networks. 
In this article, we consider an IRS-aided downlink multiple-input single-output (MISO) network, where the base station (BS) is not required to know the underlying channel distribution. 
We formulate an outage probability minimization problem by jointly optimizing the beamforming vector at the BS and the phase-shift matrix at the IRS, while taking into account the transmit power and unimodular constraints.
The formulated problem turns out to be a non-convex non-smooth stochastic optimization problem. 
To this end, we employ the sigmoid function as the surrogate to tackle the non-smoothness of the objective function. 
In addition, we propose a data-driven efficient  alternating stochastic gradient descent (SGD) algorithm to solve the problem by utilizing the historical channel samples. 
Simulation results demonstrate the performance gains of the proposed algorithm over the benchmark methods in terms of minimizing the outage probability. 
%Simulation results demonstrate that the deployment of IRS significantly reduce the outage  brought by the IRS in terms of the outage probability.
\end{abstract}

\section{Introduction}
% intro

Intelligent reflecting surface (IRS), as an emerging cost-effective technology, has the potential to achieve the concept of ``smart radio environments", and thereby significantly improving the energy efficiency and the spectrum efficiency of wireless networks \cite{zhang}. 
The IRS refers to an artificial planar surface that is composed of many low-cost passive reflecting elements.  
Each reflecting element can be independently controlled to reflect the incident signal by introducing a desired phase shift, aiming to reconfigure the propagation environment. 
In addition, IRS also has the advantages of full-band response and flexible deployment\cite{full}.

%Recently, a great deal of works in IRS have sprung up.
%It has been considered as a revolutionizing technology which is able to significantly improve the performance of the communication system by smartly reconfiguring the wireless propagation environment with the use of massive low-cost passive reflecting elements integrated on a surface\cite{zhang}.  
%It enables the transmission environment controllable.
 
% work way and structure

%In terms of the structure and work way. IRS is equipped with passive reflecting elements which are reconfigurable by a controller.
%Receiver can get the desired amplitude and phase of the reflected signals by adjust the IRS setting.
%The reconfigurable passive elements in IRS can independently reflecting the incident signals either constructively or destructively.

% related work of IRS

Due to its unique  advantages, IRS has recently attracted considerable attentions\cite{wu2019intelligent,fu2019reconfigurable,huang1,huang2019reconfigurable,channel_capacity,hua2019reconfigurable,jiang2019over}. 
In particular, the joint active and passive beamforming design was studied in IRS-aided multiple-input single-output (MISO) \cite{wu2019intelligent} and non-orthogonal multiple access (NOMA) \cite{fu2019reconfigurable,huang1} networks to minimize the transmit power consumption. 
IRS was also leveraged to enhance the energy efficiency \cite{huang2019reconfigurable}, maximize the channel capacity \cite{channel_capacity}, facilitate the edge inference \cite{hua2019reconfigurable}, as well as boost the over-the-air computation \cite{jiang2019over}. 
However, all the aforementioned studies assumed that the instantaneous channel state information (CSI) is perfectly known at the base station (BS). 

%\cite{channel_capacity} exhibits the improvement of the channel capacity brought by IRS.
%\cite{privacy} discuss that IRS can enhance the privacy security by suppress the received signal.
%In \cite{wubeamse,beamzhangdiscrete,di2019hybrid}, sum rate and transmitting power based beamforming design have been considered in IRS-aided MISO system.
%According to \cite{wubeamse}, IRS enable the spectral efficience get improved in a large degree in downlink multi-user and single-user MISO system.
%Besides,\cite{beamzhangdiscrete,di2019hybrid} have shown the discrete phase-shift design in the IRS-aided wireless communication system.

%%%%outage

% our motivation and work

%Due the passive nature of the IRS, it is generally difficult to acquire the accurate downlink CSI at the BS in practical cellular networks.
It is generally difficult to acquire the accurate downlink CSI at the BS in practical cellular networks\cite{imperfectcsi}. 
With imperfect CSI, the robust beamforming design is commonly adopted in the literature and takes into account the channel uncertainty\cite{uncertain}. 
Very recently, the authors in \cite{robust} modeled the channel estimation error as a random variable with a bounded region and developed a robust beamforming design for IRS-aided MISO networks. 
However, by ensuring the worst-case performance over the channel uncertainty region, such a robust beamforming design is very conservative and results in a poor performance \cite{shi}, which motivates this work. 

%But existing works on IRS-aided beamforming design are mostly relied on the assumption that instantaneous CSI are  available \cite{channel_capacity,privacy, beamzhangdiscrete, wubeamse,di2019hybrid}. 
%Inevitably there will be uncertainty in available CSI in practice\cite{shi}.
%\cite{robust} consider IRS-aided robust beamforming design in imperfect channel.
%It guarantees the worst case performance over the whole uncertain set.
%The primary advantage of robust formulation is the computational tractability\cite{shi}.
%However the worst-case formulation might be over-conservative\cite{shi}.   

In this article, we consider an IRS-aided MISO network, where the BS has no prior knowledge on the underlying channel distribution. 
We propose a data-driven approach that relies on a collection of channel samples to jointly optimize the beamforming vector at the BS and the phase-shift matrix at the IRS. 
The goal is to minimize the outage probability, defined as the probability that the received signal-to-noise ratio (SNR) falls below a certain threshold, while taking into account the transmit power constraint at the BS and the unimodular constraint of phase shift at the IRS. 
The formulated problem turns out to be a highly intractable non-convex non-smooth stochastic optimization problem. 
To tackle the non-smoothness of the objective function, we adopt the sigmoid function as the surrogate, which leads to a continuous optimization problem. 
To decouple the optimization variables and also reduce the computation complexity, we propose an efficient alternating stochastic gradient descent (SGD) algorithm to solve the problem. 
Simulation results demonstrate the effectiveness of the proposed algorithm and the importance of deploying an IRS in reducing the outage probability.

\section{System Model and Problem Formulation}
\subsection{System Model}
Consider an IRS-aided MISO system consisting of an $M$-antenna BS, a single-antenna user, and an IRS. 
The IRS is equipped with $N$ passive reflecting elements, each of which can be software-controlled to induce a desired phase shift on the incident signal. 
We denote $ \bm w\in\mathbb{C}^{M} $ and $ s\in\mathbb{C}$ as the beamforming vector and the information symbol at the BS,  respectively.
Without loss of generality, we assume that symbol $s$ has a zero mean and unit power. 
We denote $ \bm h_d\in\mathbb{C}^M, \bm G\in\mathbb{C}^{N\times M}$, and $\bm h_r\in\mathbb{C}^{N} $ as the channel responses of the BS-user link, BS-IRS link, and IRS-user link, respectively. 
Thus, the signal, propagating through both the direct and reflect links, received at the user can be expressed as 
\begin{align} \label{receive}
y = (\bm h_r^{\sf{H}}\bm \Theta\bm G+\bm h_d^{\sf H})\bm ws+z,
\end{align}
where $\bm\Theta=\text{diag}(\alpha \mathrm{e}^{j\theta_1}, \ldots,\alpha \mathrm{e}^{j\theta_N})$ denotes the diagonal phase-shift matrix of the IRS with $\alpha \in [0, 1]$ and $\theta_n \in [0, 2\pi)$ being the amplitude reflection coefficient and the phase shift of the $n$-th reflecting element, respectively, and $z$ denotes the additive white Gaussian noise (AWGN) with zero mean and variance $\sigma^2$. 
As in \cite{wu2019intelligent,fu2019reconfigurable,huang1,huang2019reconfigurable,channel_capacity,hua2019reconfigurable,jiang2019over}, we assume that the amplitude reflection coefficient $\alpha$ equals to one. 
Because of the high path loss, we assume that the signals after being reflected two or more times have negligible power \cite{wu2019intelligent, fu2019reconfigurable,huang1,huang2019reconfigurable,channel_capacity,hua2019reconfigurable,jiang2019over}.

According to (\ref{receive}), the signal-to-noise ratio (SNR) achieved by the user can be expressed as
\begin{align}
\Gamma = \frac{|(\bm h_r^{\sf{H}}\bm \Theta\bm G+\bm h_d^{\sf H})\bm w|^2}{\sigma^2}. 
\end{align}

\subsection{Problem Formulation}

If the instantaneous CSI is available at the BS, a typical problem formulation would be the joint design of beamforming vector $\bm w$ and phase-shift matrix $\bm\Theta$ for SNR maximization \cite{wu2019intelligent}. 
However, the instantaneous CSI is difficult, if not impossible, to obtain in practice\cite{xia}. 
To account for this situation, we consider a probabilistic model, where only a set of channel samples corresponding to an unknown underlying channel distribution are available at the BS. 
Specifically, we formulate an optimization problem to minimize the probability that the received SNR (i.e., $\Gamma$) falls below a certain threshold, denoted as $\gamma$, by jointly optimizing beamforming vector $\bm w$ and phase-shift matrix $\bm\Theta$. 
By taking into account the transmit power constraint and the unit modulus constraint, the formulated outage minimization problem can be expressed as 
\begin{align}\label{p1}
\mathop{\text{minimize}}_{ \bm w,\bm v}
&\quad \operatorname{Pr}\left(|(\bm v^{\sf H}\text{diag}(\bm h_r^{\sf{H}})\bm G+\bm h_d^{\sf H})\bm w|^2<\gamma\sigma^2\right)\nonumber \\
\text{subject to}& \quad \|\bm{w}\|^2\le P, \\
&\quad |v_n|=1, \quad n=1,\ldots,N, \nonumber
\end{align}
where $\bm v =[\mathrm{e}^{j\theta_1},\ldots, \mathrm{e}^{j\theta_N}]^{\sf H}$
and $P$ is the maximum transmit power of the BS. 
For notational ease, we define
\begin{align}\label{obj}
~d(\bm w,\bm v;\bm h_e) :=\sigma^2-\frac{1}{\gamma} \left|(\bm v^{\sf H}\text{diag}(\bm h_r^{\sf{H}})\bm G+\bm h_d^{\sf H})\bm w \right|^2, 
\end{align}
where $\bm h_e = \{\bm h_d,\bm h_r,\bm G\}$ is an abstraction of channel responses $\bm h_d$, $\bm h_r$, and $\bm G$. 
We can rewrite the outage event as
\begin{align} %\label{equi}
|(\bm v^{\sf H}\text{diag}(\bm h_r^{\sf{H}})\bm G+\bm h_d^{\sf H})\bm w|^2<\gamma\sigma^2~\Longleftrightarrow
~d(\bm w,\bm v;\bm h_e) >0. \nonumber
\end{align}
The outage probability minimization can be approximated by the maximization of the time proportion that the channel condition satisfies a target SNR requirement. 
Hence, we can rewrite problem \eqref{p1} as the following stochastic optimization problem
\begin{align}\label{p2}
\mathop{\text{minimize}}_{ \bm w,\bm v}&\quad f(\bm w,\bm v):=\mathbb{E}_{\bm h_e}\left[\mathcal{I}_{(0,+\infty)}\left(d(\bm w,\bm v;\bm h_e)\right)\right] \nonumber \\
\text{subject to} &\quad  \|\bm w\|^2\le P, \nonumber\\
&\quad |v_n|=1, \quad n=1,\ldots,N, 
\end{align}
where $\mathcal{I}_{(0,+\infty)}(x) $ is an indicator function defined as
\begin{align}\label{indicator}
\mathcal{I}_{(0,+\infty)} (x)=\begin{cases}
1,\quad\text{if} \; x>0, \\
0,\quad\text{otherwise}. 
\end{cases}
\end{align}

Without prior knowledge on the underlying channel distribution, we adopt a data-driven approach that relies on a collection of channel samples to optimize the beamforming vector at the BS and the phase-shift vector at the IRS for outage minimization. 
In particular, we denote the set of channel samples available at the BS as $\mathcal{H}_T = \{ \bm h_e^t \}_{t=1}^{T}$, where $\bm h_e^t$ denotes the $t$-th channel sample. 
Note that these channel samples can be obtained via measurement within a certain time period, as discussed in \cite{shiyunmei}. 
With the set of channel samples, we adopt the following sample average approach to approximate $f(\bm w, \bm v)$
\begin{align}
\hat{f}(\bm w, \bm v; \mathcal{H}_T) := \frac{1}{T} \sum_{t=1}^{T} \mathcal{I}_{(0,+\infty)}\left(d(\bm w,\bm v;\bm h_e^t)\right). 
\end{align}

It is worth noting that with the ergodicity of the channel process, the sample average converges to the ensemble average with probability $1$ as $T \to \infty$\cite{saa}. 
As a result, we obtain the following optimization problem for solving problem \eqref{p2}
\begin{align}\label{ps}
\mathop{\text{minimize}}_{ \bm w,\bm v}&\quad \hat{f}(\bm w, \bm v; \mathcal{H}_T) \nonumber \\
\text{subject to} &\quad  \|\bm w\|^2\le P, \\
&\quad |v_n|=1, \quad n=1,\ldots,N. \nonumber
\end{align}

Problem $\eqref{ps}$ is difficult to be solved due to the following three challenges.
First, the objective function $\hat{f}(\bm w,\bm v; \mathcal{H}_T)$ is non-convex and discontinuous.
Second, the optimization variables $\bm w$ and $\bm v$ are coupled in the objective function. 
Third, the unimodular constraint of phase shift vector is also non-convex. 
In the following section, we shall propose an alternating SGD algorithm to solve this problem. 

\section{Alternating SGD Optimization Framework}
In this section, we first present a smooth surrogate for the indicator function and then propose an alternating SGD optimization algorithm to solve problem \eqref{p3}, where the beamforming vector at the BS and the phase-shift vector at the IRS are alternatively optimized until convergence. 
\subsection{Smooth Surrogate for Indicator Function}
To handle the non-convexity and discontinuity of the objective function, we adopt the continuous and smooth sigmoid function as the surrogate of the indicator function $\mathcal{I}_{(0,+\infty)}(x)$ \cite{shiyunmei}. 
In particular, the sigmoid function is defined as  
\begin{align}\label{sigmoid}
	\mathcal{S}(z) = \frac{1}{1+e^{-z}}. 
\end{align}
Note that $\mathcal{S}(z) \rightarrow 1$ when $z \rightarrow \infty$ and $\mathcal{S}(z) \rightarrow 0$ when $z \rightarrow -\infty$. 
Although the sigmoid function is not convex, it is continuously differentiable and strictly monotonic increasing. 
To this end, problem $\eqref{ps}$ can be rewritten as 
\begin{align}\label{p3}
\mathop{\text{minimize}}_{\bm{w},\bm v}&\quad f_1(\bm w,\bm v; \mathcal{H}_T) := \frac{1}{T} \sum_{t=1}^{T} S\left(d(\bm w,\bm v;\bm h_e^t)\right)
\nonumber \\
\text{subject to}&\quad
\|\bm w\|^2\le P, \nonumber\\
&\quad |v_n|=1, \quad n=1,\ldots,N. 
\end{align}

After replacing the indicator function with the sigmoid function, 
problem $\eqref{p3}$ becomes a continuous optimization problem. 
With a differentiable objective function, the gradient descent (GD) method can be adopted to solve the problem. 
To obtain the gradient of $f_1(\bm w,\bm v; \mathcal{H}_T)$, we need to compute the gradients of $\mathcal{S}\big(d(\bm w, \bm v;\bm h_e^t)\big)$ for all channel samples in each iteration. 
However, the computation complexity increases significantly as the number of channel samples increases. 
To reduce the computation complexity, we adopt the SGD method to solve problem $\eqref{p3}$. 
 Specifically, we randomly choose a channel sample from set $\mathcal{H}_T$ to compute the gradient of $\mathcal{S}\big(d(\bm w, \bm v;\bm h_e^t)\big)$. 
To decouple the optimization variables, we propose an alternating SGD algorithm in the following.

\subsection{Beamforming Vector Optimization} \label{subsecb}
When the phase-shift vector $ \bm v $ is fixed, the cascaded channel response, denoted as $\bm h^{\sf H}= \bm v^{\sf H}\text{diag}(\bm h_r^{\sf{H}})\bm G+\bm h_d^{\sf H}$, is also fixed. 
We can rewrite $d(\bm w,\bm v;\bm h_e)$ as $d(\bm w;\bm h) = \sigma^2-\frac{1}{\gamma} |\bm h^{\sf{H}}\bm w|^2$.
As a result, problem \eqref{p3} can be rewritten as 
\begin{align}\label{p4}
\mathop{\text{minimize}}_{\bm w}&\quad \frac{1}{T} \sum_{t=1}^{T} S\left(d(\bm w;\bm h^t)\right)\nonumber\\
\text{subject to}&\quad \|\bm w\|^2\le P, 
\end{align}
where $\bm h^t$ is obtained based on the $t$-th channel sample. 

It is worth noting that $d(\bm w;\bm h^t)$ is a real function of complex variables. 
To facilitate the gradient calculation, we define 
\begin{align}
\tilde{\bm w}  &:=[\Re[\bm w]^{\sf T}, \Im[\bm w]^{\sf T}]^{\sf T}\in \mathbb{R}^{2M}, \\
\tilde{\bm H}  &:= \begin{bmatrix}
\Re[\bm h] & -\Im[\bm h] \\
\Im[\bm h] & \Re[\bm h]
\end{bmatrix}\in \mathbb{R}^{2M\times2}, 
\end{align}

\noindent where $\Re[\bm \cdot]$ and $\Im[\bm \cdot]$ represent the real and imaginary part of an element, respectively.
In order to explicitly exhibit parameter dependency, 
we denote $d_1(\tilde{\bm w}; \tilde{\bm H}) \hspace{-6 mm} \qquad= \sigma^2-\frac{1}{\gamma} \|\tilde{\bm H}^{\sf T}\tilde{\bm w}\|^2$.
We can equivalently express $\eqref{p4}$ in terms of real variables as
\begin{align} \label{OP_w}
\mathop{\text{minimize}}_{\tilde{\bm w}}&\quad u(\tilde{\bm w}):=\frac{1}{T} \sum_{t=1}^{T} \mathcal{S}\big(d_1(\tilde{\bm w}; \tilde{\bm H}^t)\big) \nonumber\\
\text{subject to}& \quad \|\tilde{\bm w}\|^2\le P, 
\end{align}
where $\tilde{\bm H}^t$ is obtained based on $\bm h^t$. 
%The following lemma presents the gradient of $\mathcal{S}\big(d_1(\tilde{\bm w}; \tilde{\bm H})\big)$ with respect to $\tilde{\bm w}$. 
\begin{lemma}\label{lamma1}
	The gradient of  $\mathcal{S}\big(d_1(\tilde{\bm w}; \tilde{\bm H})\big)$ denoted as $\nabla_{\tilde{\bm w}}\mathcal{S}\big(d_1(\tilde{\bm w}; \tilde{\bm H})\big) \hspace{-6mm} \qquad$ with respect to $\tilde{\bm w}$ is given by
	\begin{align}
	%\nabla_{\tilde{\bm w}}\mathcal{S}\big(d_1(\tilde{\bm w}; \tilde{\bm H})\big) \hspace{-6mm} &\qquad = 
	\mathcal{S}\big(d_1(\tilde{\bm w}; \tilde{\bm H})\big)\big(1-\mathcal{S}\big(d_1(\tilde{\bm w}; \tilde{\bm H})\big)\big)
	\big(- \frac{2}{\gamma}\tilde{\bm H} \tilde{\bm H}^{\sf T}\tilde{\bm w}\big).\nonumber
	\end{align}
\end{lemma}
%\noindent We minimize $u(\tilde{\bm w})$ by using the SGD update $\bm {\tilde w}$. 
\noindent We update $\tilde {\bm w}$ by SGD to minimize $u(\tilde{\bm w})$.
Specifically, we randomly choose a channel sample $\tilde{\bm H}^i$ to compute the gradient instead of averaging the gradient over all channel samples.
Hence, we update the beamforming vector at the BS as follows
\begin{align} \label{update_w}
\tilde{\bm y}^{k+1} &= \tilde{\bm w}^{k} - l_{\tilde{w}}\nabla_{\tilde{\bm w}}\mathcal{S}\big(d_1(\tilde{\bm w}; \tilde{\bm H^i})\big)\Big|_{\tilde{\bm w}^k},  \\ 
%\bm{\tilde{w}^{k+1}} &= \bm{Proj}\left(\bm{\tilde{y}^{k+1}}\right), \label{solve_w}\\
\tilde{\bm w}^{k+1}&=\begin{cases}
{P^{\frac{1}{2}}\frac{\tilde{\bm y}^{k+1}}{\Vert{\tilde{\bm y}^{k+1}\Vert}}},\quad\text{if} \; \Vert \tilde{\bm y}^{k+1}\Vert^2 \geq P, \\
\quad \tilde{\bm y}^{k+1},\qquad\text{otherwise}, 
\end{cases}\label{solve_w}
\end{align}
where $l_{\tilde{w}}$ denotes the step size.
To account of the power constraint (i.e., $\Vert \tilde{\bm w} \Vert^2 \leq P $), we take the Euclidean projection on $\tilde{\bm y}^{k+1}$ to obtain $\tilde{\bm w}^{k+1}$ as \eqref{solve_w}.

\subsection{Phase-Shift Vector Optimization} \label{subsecc}
With a given beamforming vector $ \bm w $, we denote $\bm a =\text{diag}(\bm h_{r}^{\sf H})\bm G\bm w \in \mathbb{C}^{N \times 1}$ and $ b = \bm h_{d}^{\sf H}\bm w \in \mathbb{ C}$ for ease of notations. 
We can rewrite $d(\bm w,\bm v; \bm h_{e}  )$ as 
%\begin{align}\label{fixm}
$d(\bm v;\bm a, b) = \sigma^2 - \frac{1}{\gamma}|b+\bm v^{\sf H}\bm a|^2$, 
%\end{align}
where $\bm v^{\sf H}\bm a = \bm v^{\sf H}\text{diag}(\bm h_{r}^{\sf H}) \bm G\bm w$. 
Hence, we can simplify problem $ \eqref{p3}$ as follows
\begin{align}\label{p6}
\mathop{\text{minimize}}_{\bm v}&\quad  \frac{1}{T} \sum_{t=1}^{T} \mathcal{S}\big(d(\bm v; \bm a^t,b^t)\big) \nonumber \\
\text{subject to}
&\quad|v_n|=1, \quad n=1,\ldots,N. 
\end{align}

\noindent To facilitate the gradient calculation, we define
\begin{align}
\tilde{\bm v}  & :=[\Re[\bm v]^{\sf T}, \Im[\bm v]^{\sf T}]^{\sf T}\in \mathbb{R}^{2N}, \\
\tilde{\bm b}  & :=[\Re[ b], \Im[  b] ]^{\sf T}\in \mathbb{R}^{2}, \\
\tilde{\bm A}  & := \begin{bmatrix}
\Re[\bm a] & -\Im[\bm a] \\
\Im[\bm a] & \Re[\bm a]
\end{bmatrix}\in \mathbb{R}^{2N\times2}. 
\end{align}

\begin{figure*}[h]
        \centering
        \subfigure[Outage probability versus number of reflecting elements at IRS when $M = 15$ and $\gamma = 3$.]{
        \label{N}
        \includegraphics[width=0.65\columnwidth]{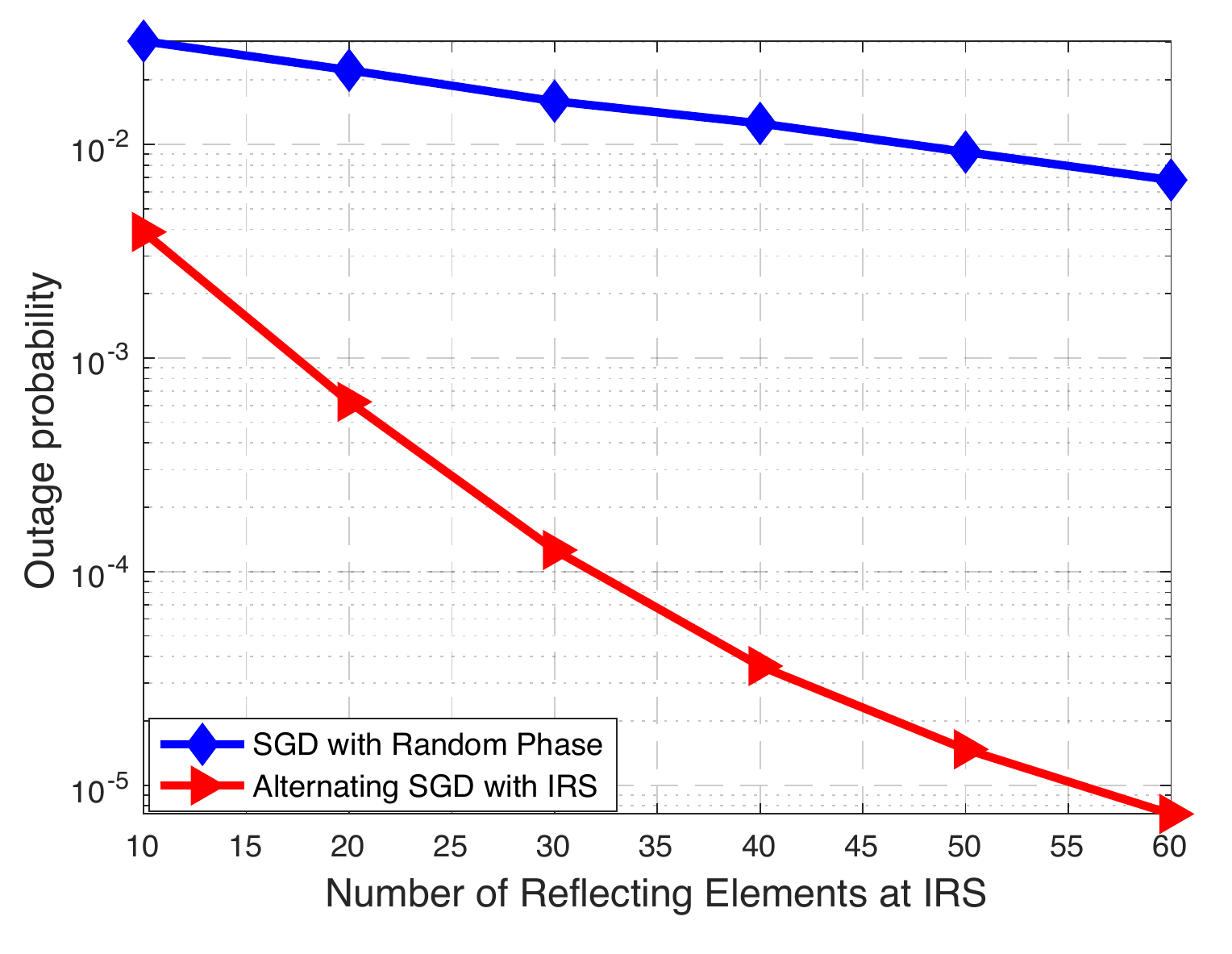}}
        \subfigure[Outage probability versus reception threshold when $M = 15$ and $N=50$.]{
        \label{gamma}
        \includegraphics[width=0.67\columnwidth]{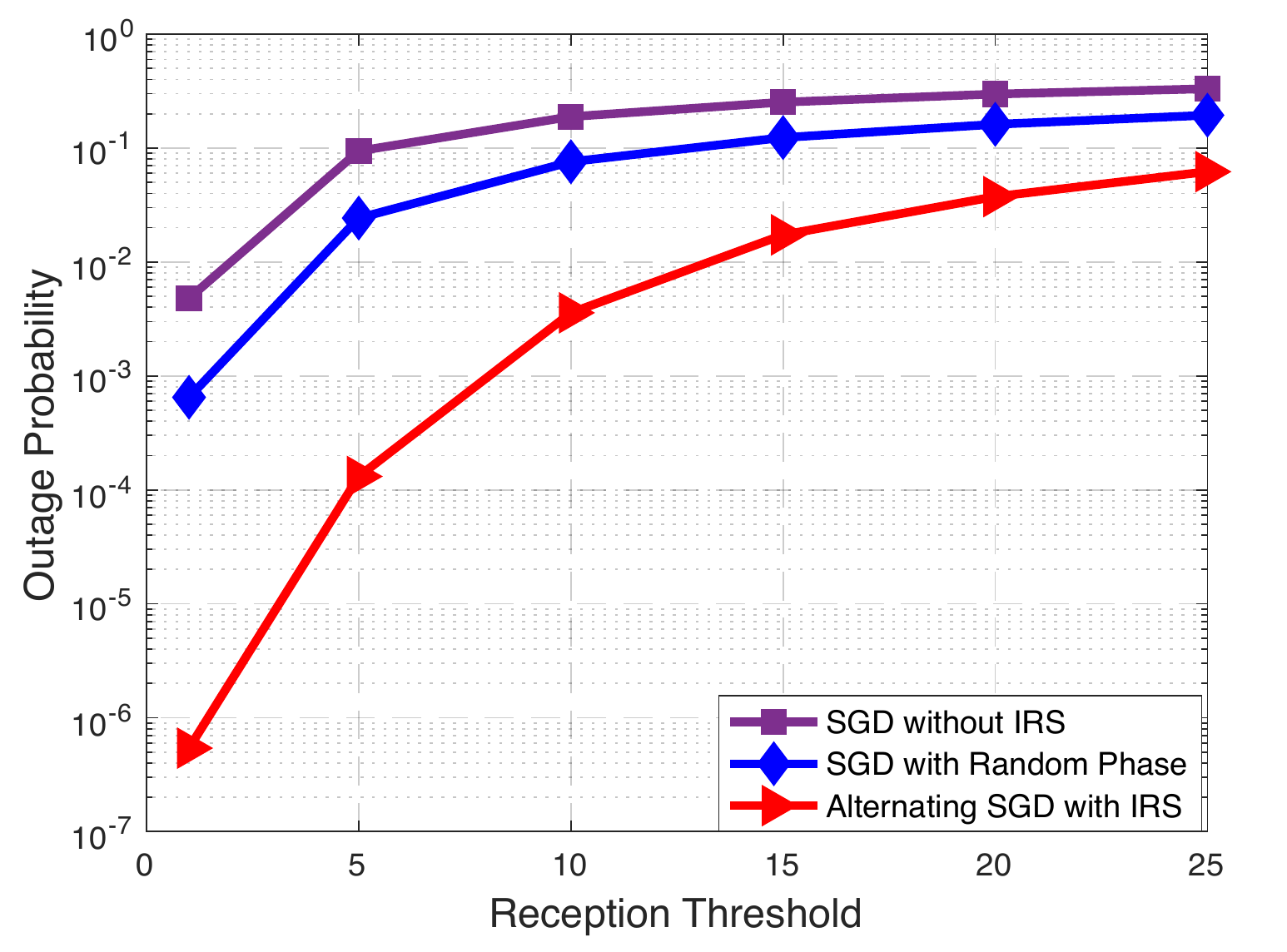}}
        \subfigure[Outage probability versus number of antennas at the BS when $\gamma = 5$ and $N=30$.]{
        \label{M}
        \includegraphics[width=0.65\columnwidth]{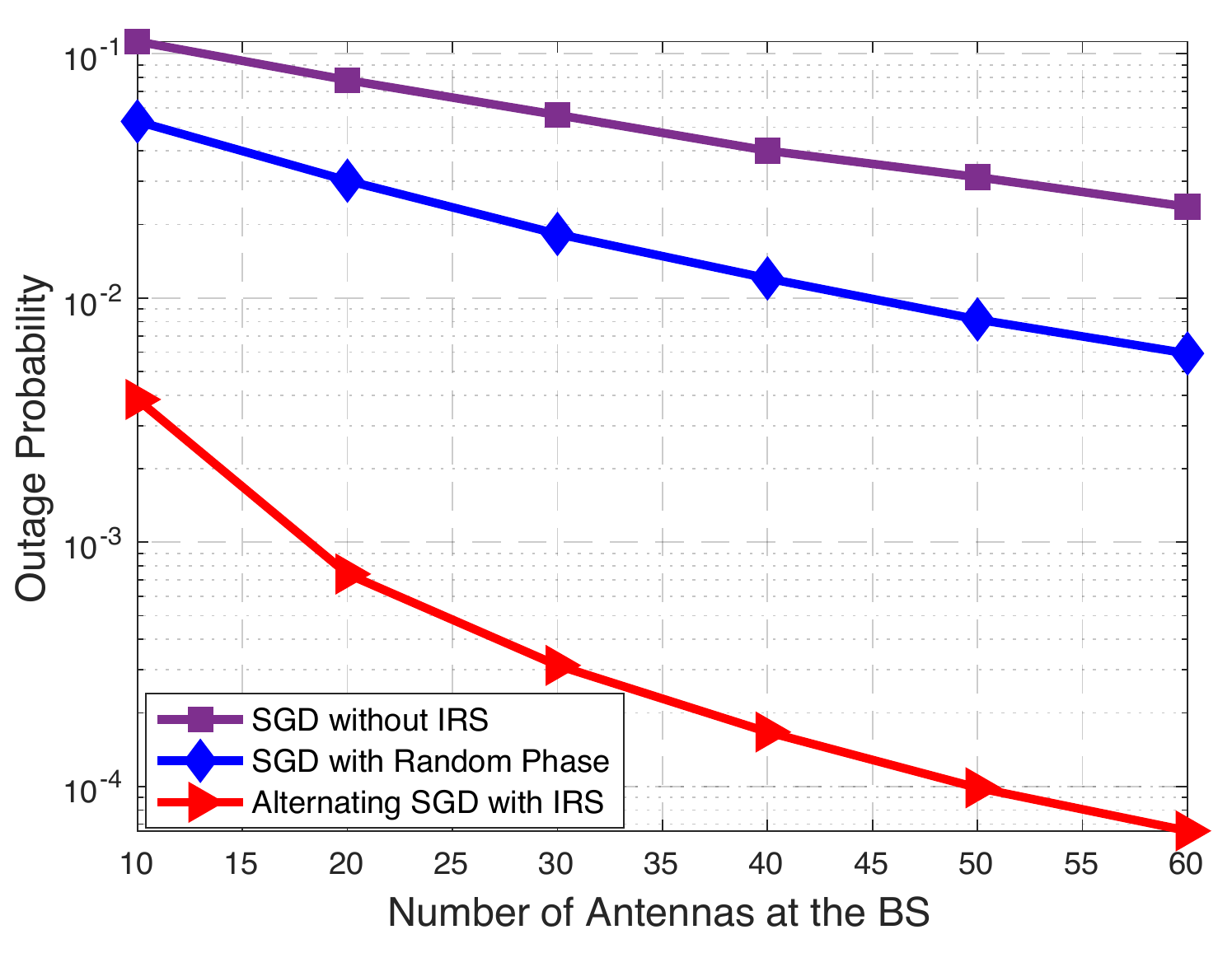}} 
        \caption{Outage probability versus different system parameters in an IRS-aided MISO wireless network.}
        \vspace{-5mm}
\end{figure*}
\noindent By denoting $d_2(\tilde{\bm v};\tilde{\bm A},\tilde b) = \sigma^2 - \frac{1}{\gamma}\| \tilde{ \bm b}+\tilde{\bm A}^{\sf T}\tilde{\bm v}\|^2$, 
we can equivalently express problem $\eqref{p6}$ in terms of real variables
\begin{align}\label{p7}
\mathop{\text{minimize}}_{\tilde{\bm v}}&\quad  g(\tilde{\bm v}):=\frac{1}{T} \sum_{t=1}^{T} \mathcal{S}\big(d_2(\tilde{\bm v};\tilde{\bm A}^t,\tilde b^t)\big) \nonumber \\
\text{subject to}&\quad |\tilde{v}_n|^2 + |\tilde{v}_{n+N}|^2=1,\quad n=1,\ldots,N.
\end{align}

%The following lemma presents the gradient of $\mathcal{S}\big(d_2(\tilde{\bm v};\tilde{\bm A},\tilde b)\big)$ with respect to $\tilde{\bm v}$. 
\begin{lemma}\label{lamma2}
	The gradient of $\mathcal{S}\big(d_2(\tilde{\bm v};\tilde{\bm A},\tilde b)\big)$ denoted as $\nabla_{\tilde{\bm v}}\mathcal{S}\big(d_2(\tilde{\bm v};\tilde{\bm A},\tilde b)\big)$ with respect to $\tilde{\bm v}$ is given by
	\begin{align}
	%\nabla_{\tilde{\bm v}}\mathcal{S}\big(d_2(\tilde{\bm v};\tilde{\bm A},\tilde b)\big) & = 
	\mathcal{S}\big(d_2(\tilde{\bm v};\tilde{\bm A},\tilde b)\big)\big(1-\mathcal{S}\big(d_2(\tilde{\bm v};\tilde{\bm A},\tilde b)\big)\big) \big(-\frac{2}{\gamma}(\tilde{\bm A}\tilde{\bm b}+\tilde{\bm A} \tilde{\bm A}^{\sf T}\tilde{\bm v})\big). \nonumber
	\end{align} 
\end{lemma}

\noindent Similarly, we update the phase-shift vector by using the SGD method as follows
\begin{align} \label{update_v}
\tilde{\bm y}^{k+1} &= \tilde{\bm v}^{k} - l_{\tilde{v}}\nabla_{\tilde{\bm v}}\mathcal{S}\big(d_2(\tilde{\bm v};\tilde{\bm A}^i,\tilde b^i)\big)\big|_{\tilde{\bm v}^k}, \\
%\bm{\tilde{v}^{k+1}} &= \bm{Proj}\left(\bm{\tilde{y}^{k+1}}\right).\label{solve_v}
\tilde{v}^{k+1}_{n} &= \frac{\tilde{y}^{k+1}_{n}}{(\vert\tilde{y}^{k+1}_{n}\vert^2+\vert\tilde{y}^{k+1}_{n+N}\vert^2)^{\frac{1}{2}}}, \quad n=1,\ldots,2N, \label{solve_v}
%\bm{\tilde{v}^{k+1}_{i}} &=\begin{cases}
%\bm{\frac{\tilde{y}^{k+1}_{i}}{(\vert\tilde{y}^{k+1}_{i}\vert^2+\vert\tilde{y}^{k+1}_{i+N}\vert^2)^{\frac{1}{2}}}},\quad\text{if} \; \quad \forall  i\in\{1,\cdots,N\}, \\
%\bm{\frac{\tilde{y}^{k+1}_{i}}{(\vert\tilde{y}^{k+1}_{i}\vert^2+\vert\tilde{y}^{k+1}_{i-N}\vert^2)^{\frac{1}{2}}}},\quad\text{if} \; \quad \forall  i\in\{1+N,\cdots,2N\}, 
%\end{cases}\label{pro}
\end{align}
where $ l_{\tilde{v}}$ denotes the step size and $i$ is randomly chosen from $\{1,\ldots,T\}$. 
Due to the unimodular constraint, we obtain $\tilde{\bm v}^{k+1}$ according to \eqref{solve_v}, where
$\tilde{y}^{k+1}_{n}=\tilde{y}^{k+1}_{n-2N}$ if $n > 2N $.
%Note that, in each iteration, GD requires the calculation of $T$ gradients, while SGD only calculates the gradient once. 

The overall algorithm for solving problem \eqref{ps} is termed as the alternating SGD algorithm and summarized in Algorithm \ref{algo1}, which alternatively solves problems \eqref{OP_w} and \eqref{p7} until convergence. 
To ensure the objective function declining smoothly and converging to a constant, the step size should be decreased as the iteration process proceeds. 
\begin{algorithm} [t]
	\caption{Proposed Alternating SGD Algorithm}
	\label{algo1}
	\SetKwData{Index}{Index}
    \KwIn{Threshold $\epsilon$, Max number of iterations $J$, $K$.}
    	\For{$j=1\rightarrow J$}{
    	%Get $T$ samples for this iteration.\\
		 \For{$k=1\rightarrow K$}{
		 	Fix $\tilde{\bm v}$, update $\tilde{\bm w}$ according to \eqref{update_w} and \eqref{solve_w}. \\
					   Calculate objective value of (\ref{OP_w}) as $O_{\tilde{\bm w}}^{k+1}$. \\
			  			   \If {$|O_{\tilde{\bm w}}^{k} - O_{\tilde{\bm w}}^{k+1}|/{O_{\tilde{\bm w}}^{k}} \leq \epsilon$}{
			   
			   break}
			 }
         \For{$k=1\rightarrow K$}{
		 	Fix $\tilde{\bm w}$, update $\bm{\tilde{v}}$ according to \eqref{update_v} and \eqref{solve_v}.\\
						  Calculate objective value of (\ref{p7}) as $O_{\bm{\tilde{v}}}^{k+1}$. \\
			  \If {$|O_{\tilde{\bm v}}^{k} - O_{\tilde{\bm v}}^{k+1}|/{O_{\tilde{\bm v}}^{k}} \leq \epsilon$}{
			  break}
             }
    %Update $l_{\tilde{w}}$ and $l_{\tilde{v}}$
   }
% \KwOut{$\bm{outage \quad probability},\bm{\tilde{w},\tilde{v}}$}
\end{algorithm}
\vspace{-3mm}
\section{Simulation Results}
In this section, we present the simulation results of the proposed alternating SGD algorithm in an IRS-aided MISO system. 
We consider a three-dimention coordinate system, where the BS and the IRS are located at $(0, 0, 10)$ and $(15, 5, 5)$ meters, respectively. 
In addition, the user is randomly located at the square centered at $(18,1,0)$ and the side length of $2$ meters.
All links suffer from path loss and Rayleigh fading. 
The path loss with length $d$ is modeled as $d^{-\beta}$, where $\beta$ denotes the path loss exponent. 
For the BS-user link, the BS-IRS link and the IRS-user link, $\beta$ is set as 2.5, 2.1 and 2.2, respectively. 
In simulations, we set the size of the channel sample set $T = 250$, and the maximum transmit power of the BS and the noise power as $30$ dBm and $-80$ dBm, respectively. 
We set the step size $l_{\tilde{w}}=1$ and $ l_{\tilde{v}}=0.1$ initially. They gradually decay with rate $0.99$ as the iteration times increasing. We set the max number of iterations $J=1000$, $K=5000$ and threshold  $\epsilon = 10^{-5}$.
 The results in each figure are obtained by averaging over $60$ Monte Carlo realizations.

Fig.\ref{N} shows the outage probability versus the number of elements at the IRS when $M = 15$ and $\gamma = 3$. 
We compare the performance of the proposed alternating SGD algorithm with a benchmark method termed as SGD with random phase, where the phase-shift vector of the IRS is randomly chosen and keeps fixed when solving the outage minimization problem. 
The outage probability decreases as the number of reflecting elements increases, because of the increased power received at the user. 
Compared to the SGD with random phase method, the proposed alternating SGD algorithm achieves a much lower outage probability. 
This demonstrates the importance of optimizing the phase shifts of the IRS.

Fig. \ref{gamma} shows the impact of outage threshold $\gamma$ on the outage probability when $M = 15$ and $N=50$. 
As the outage threshold increases, the outage probability also increases. 
Moreover, the proposed algorithm outperforms the SGD with random phase and the SGD without IRS methods in terms of the outage probability. 
This demonstrates the importance of deploying an IRS in enhancing the transmission reliability.

Fig. \ref{M} shows the impact of the antenna number at the BS on the outage probability when $\gamma = 5$ and $N=30$. 
The outage probabilities of all methods decrease as the number of antennas at the BS increases. 
This is due to the fact a higher power gain can be achieved with a larger number of antennas. 
Moreover, the proposed alternating SGD algorithm achieves a much better performance than the two benchmark methods for different number of BS antennas.

\section{Conclusions}
In this article, we investigated the outage probability minimization problem for an IRS-aided MISO network, under  
the condition that the underlying channel distribution is not available at the BS. 
We proposed the joint design of the beamforming vector at the BS and the phase-shift matrix at the IRS to minimize the outage probability, taking into account the transmit power constraint and the unimodular constraint. 
%We transformed the formulated stochastic optimization problem to a sample average minimization problem. 
We transformed the formulated problem to a stochastic optimization problem. 
The sigmoid function was adopted to tackle the non-smoothness of the objective function and the sample average approach was applied to approximate the expectation. 
Furthermore, we proposed a low-complexity alternating SGD algorithm to solve the problem. 
%Simulation results showed that the proposed algorithm outperforms the state-of-the-art methods in terms of the outage probability. 
%Simulation results showed that the proposed algorithm performs well in this stochastic optimization problem and the considerable performance gains achieved by the IRS in terms of minimizing the outage probability.
Simulation results demonstrated the effectiveness of the proposed algorithm.\bibliographystyle{IEEEtran}
\bibliography{Minimum_Outage} 
\end{document}